\documentclass[preprint,showkeys,preprintnumbers,amsmath,amssymb]{revtex4}  

\usepackage{graphicx}
\usepackage{dcolumn}
\usepackage{bm}
\usepackage{multirow}
\usepackage{color}
\usepackage{hyperref}
\usepackage[normalem]{ulem}

\begin{document}

\title{Valley dynamics probed through charged and neutral exciton emission in monolayer WSe$_2$}

\author{G. Wang}
\author{L. Bouet}
\author{D. Lagarde}
\author{M. Vidal}
\author{A. Balocchi}
\author{T. Amand}
\author{X. Marie}
\author{B. Urbaszek}
\email[]{urbaszek@insa-toulouse.fr}
\affiliation{%
Universit\'e de Toulouse, INSA-CNRS-UPS, LPCNO, 135 Av. de Rangueil, 31077 Toulouse, France}


\begin{abstract}
Optical interband transitions in monolayer transition metal dichalcogenides such as WSe$_2$ and MoS$_2$ are governed by chiral selection rules. This allows efficient optical initialization of an electron in a specific K-valley in momentum space. Here we probe the valley dynamics in monolayer WSe$_2$ by monitoring the emission and polarization dynamics of the well separated neutral excitons (bound electron hole pairs) and charged excitons (trions) in photoluminescence. The neutral exciton photoluminescence intensity decay time is about 4ps, whereas the trion emission occurs over several tens of ps. The trion polarization dynamics shows a partial, fast initial decay within tens of ps before reaching a stable polarization of  $\approx$ 20\%, for which a typical valley polarization decay time larger than 1ns can be inferred. This is a clear signature of stable, optically initialized valley polarization. 

\end{abstract}

                          \keywords{WSe2, valley dynamics, time resolved photoluminescence, transition metal dichalcogenides, two dimensional materials} 
\maketitle

In strong analogy to graphene, the physical properties of transition metal dichalcogenides (TMDCs) change drastically when thinning the bulk material down to one monolayer (ML) \cite{Geim:2013a}. The closely related ML materials WSe$_2$, MoS$_2$, MoSe$_2$ and WS$_2$ \cite{Zhu:2011a} have a direct bandgap in the visible region \cite{Mak:2010a,Splendiani:2010a,Zhao:2013b} and show strong optical absorption. ML WSe$_2$ is an exciting, atomically flat, two-dimensional material for electronics \cite{Fang:2012a,Wei:2013a}, non-linear optics \cite{Zeng:2013a} and optoelectronics \cite{Ross:2013a,Sundaram:2013a}, just as ML MoS$_2$ \cite{Radisavljevic:2011a,Sundaram:2013a,Kumar:2013a}.
Current micro-and nano-electronics is based on the manipulation of the electron charge and spin. ML TMDCs provide unique and convenient access to controlling in addition the electron valley degree of freedom in k-space in the emerging field of 'valleytronics' \cite{Xiao:2012a,Rycerz:2007a,Zhu:2012a}. 
\begin{figure}
\includegraphics[width=0.47\textwidth]{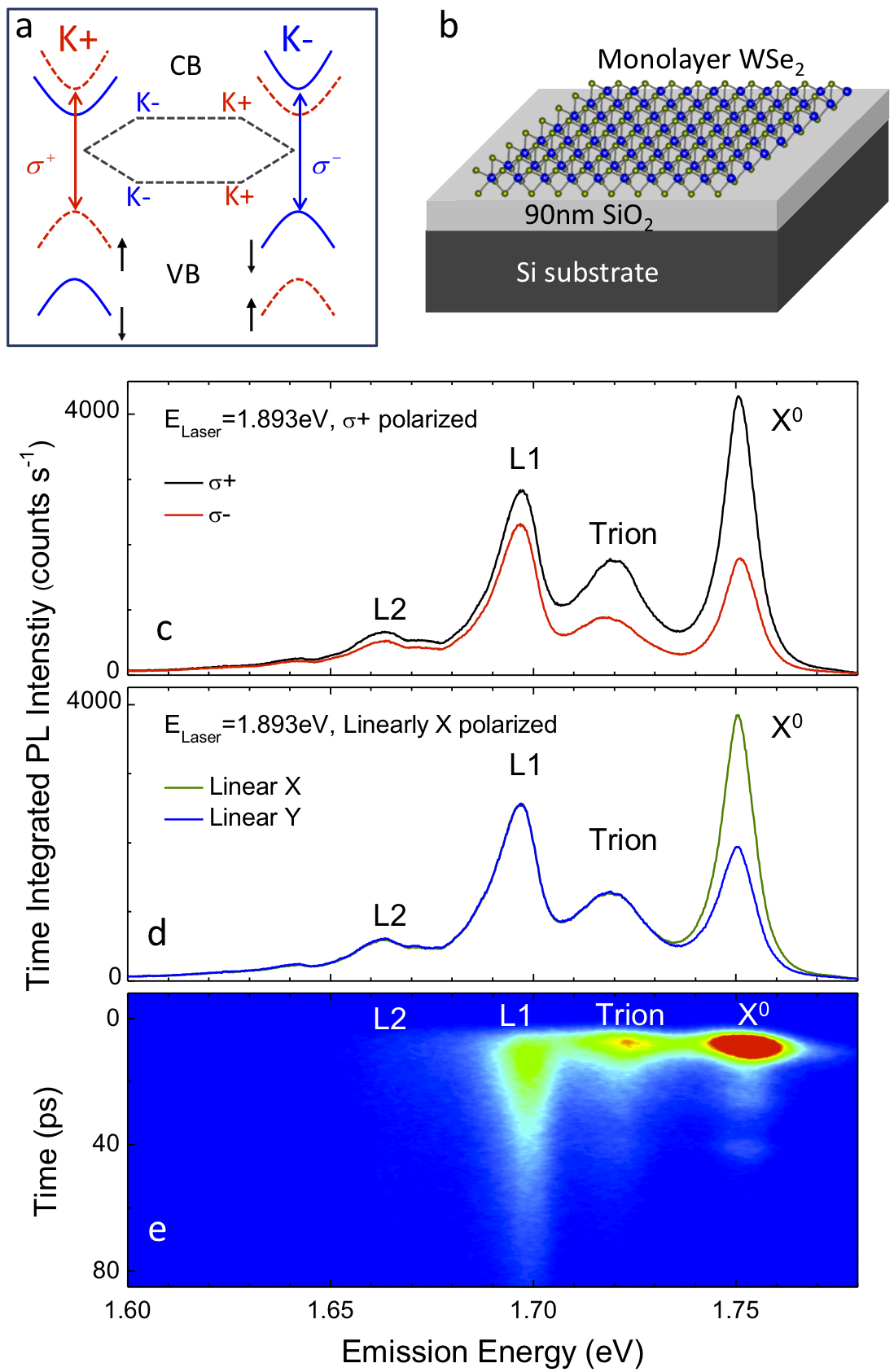}
\caption{\label{fig:fig1}  (a) optical interband selection rules for monolayer WSe$_2$ according to \cite{Liu:2013a} (b) investigated sample structure (c) Laser polarization $\sigma^+$ and E$_\text{Laser}=1.893$~eV. PL Emission of ML WSe$_2$ at T=4K, the X$^0$, trion and localized states are marked. black (red) $\sigma^+(\sigma-)$ polarized   (d) Linear Laser polarization X. green (blue) corresponds to linear X (linear Y) polarized emission. (e) Streak camera image of TRPL total intensity showing different emission times for X$^0$, trion and localized states. Blue (red) corresponds to zero (10000) counts.}
\end{figure} 
The circular polarization ($\sigma^+$ or $\sigma^-$) of the absorbed or emitted photon can be directly associated with selective carrier excitation in one of the two non-equivalent K valleys ($K_+$ or $K_-$, respectively) \cite{Zhu:2011a,Cao:2012a,Xiao:2012a,Xiao:2013a}, see figure \ref{fig:fig1}a. 
The valley polarization is protected by the strong spin-orbit splitting in the valence and conduction band \cite{Xiao:2012a,Liu:2013a,Kosmider:2013a}, leading in principle to a high stability for the valley degree of freedom. The high circular photoluminescence (PL) polarization degree reported in time integrated measurements in ML MoS$_2$ \cite{Jones:2013a,Cao:2012a,Mak:2012a,Sallen:2012a,Kioseoglou:2012a,Wu:2013a} and ML WSe$_2$ \cite{Jones:2013a} seems to confirm this prediction. \\
\indent The stability of the created valley polarization is crucial for manipulating the electron valley degree of freedom in transport measurements or with successive laser pulses in optical control schemes, where excitonic effects are important \cite{Qiu:2013a,Ramasubramaniam:2012a}. Recent time resolved studies show PL  emission times in the picosecond range \cite{Lagarde:2014a,Korn:2011a} and pump-probe measurements in ML MoS$_2$ have shown polarization decay times in the ps range \cite{Mai:2014a,Wang:2013d} corresponding to fast relaxation of the valley index.  In these experiments on ML MoS$_2$ the neutral exciton (X$^0$) and the charged exciton (trion) emission cannot be clearly spectrally separated due to the broad transitions, although the evolution of the valley polarization is expected to be distinctly different for the two complexes. The neutral excitons in different K-valleys are coupled by Coulomb exchange \cite{Maialle:1993a},  which can lead to inter-valley scattering \cite{Yu:2014a,Yu:2014b}. The trion polarization is expected to be far more stable as inter-valley scattering demands in this case spin-flips of individual carriers, which are energetically and spin-forbidden. This would make the trion an excellent candidate for optically initialized valley Hall experiments \cite{Yu:2014b}. However, the trion valley dynamics in ML TMDCs is so far unexplored.\\
\indent In time resolved PL (TRPL) experiments we uncover marked differences between the X$^0$ and the trion valley dynamics in ML WSe$_2$ as the spectrally well-separated transition can be analysed independently at low temperature, see figure \ref{fig:fig1}e.  We measure a trion emission time of $\approx18~$ps. Following optical initialization with a circularly polarized laser, the trion PL emission reaches a stable polarization within about 12ps. For the strong, remaining polarization we can infer a decay time longer than 1ns. This is a direct experimental signature of the temporal stability of optically generated valley polarization. In contrast, the neutral exciton emission and polarization decays within a few ps. We also clearly identify localized excitons via their characteristically long emission times due to the lower dipole oscillator strength.

\begin{figure}
\includegraphics[width=0.47\textwidth]{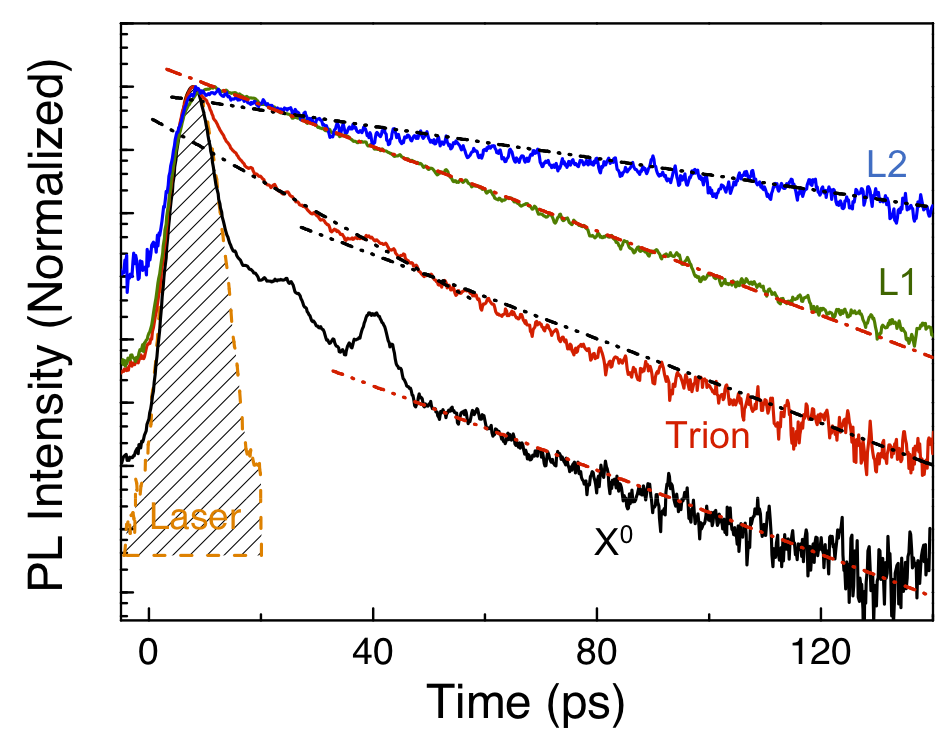}
\caption{\label{fig:fig2} \textbf{Time resolved photoluminescence.}  T=4K. Normalized PL dynamics (total peak intensity) in log-scale for X$^0$, trion, L1 and L2. The lines show fits with exponential decays with typical times of 30~ps (X$^0$ and L1) and 78~ps (L2).  Trion fitted with bi-exponential decay, with 18~ps and 30~ps characteristic decay times. The small peaks superimposed on the X$^0$ and trion dynamics come from laser reflections.}
\end{figure} 

\textbf{Samples and Experimental Set-up.---} Monolayer WSe$_2$ flakes are obtained by micro-mechanical cleavage of a bulk WSe$_2$ crystal \cite{Novoselov:2005a} (from 2D Semiconductors, USA) on 90 nm SiO$_2$ on a Si substrate, see figure \ref{fig:fig1}b. The 1ML region is identified by optical contrast and very clearly in PL spectroscopy. Experiments between T=4 and 300K are carried out in a confocal microscope optimized for polarized PL experiments \cite{Sallen:2011a}.  The WSe$_2$ flake is excited by picosecond pulses generated by a tunable frequency-doubled optical parametric oscillator (OPO) synchronously pumped by a mode-locked Ti:Sa laser. The typical pulse and spectral width are 1.6 ps and 3 meV  respectively; the repetition rate is 80 MHz. The laser power has been kept in the $\mu W$ range in the linear absorption regime (see figure \ref{fig:figS2}). The laser wavelength can be tuned between 500 and 740 nm. The detection spot diameter is $\approx1$~$\mu$m. For time-integrated experiments, the PL emission is dispersed in a spectrometer and detected with a Si-CCD camera. For time-resolved experiments, the PL signal is dispersed by an imaging spectrometer and detected by a synchro-scan Hamamatsu Streak Camera  with an overall time resolution of 4 ps. The PL polarization $P_c$ defined as $P_c = \frac{I_{\sigma+}-I_{\sigma-}}{I_{\sigma+}+I_{\sigma-}}$ is analyzed by a quarter-wave plate placed in front of a linear polarizer. Here $I_{\sigma+}(I_{\sigma-})$ denotes the intensity of the right ($\sigma^+$) and left $(\sigma^-)$ circularly polarized emission. \\

\textbf{Experimental Results.---} For PL experiments the laser excitation energy is E$_\text{Laser}=1.893$~eV, which is 140~meV above the neutral A-exciton emission energy and clearly below the B-exciton absorption as confirmed in reflectivity measurements (see figure \ref{fig:figS1}). The time integrated PL emission at T=4K of the WSe$_2$ monolayer stems from the recombination of X$^0$, trions and localised excitons. These lines are very similar to the emission reported for this system in \cite{Jones:2013a}, where a bias was applied to the ML WSe$_2$. At this point it is unclear whether the trion emission stems from positively or negatively charged excitons. Considering the commonly observed residual n-type doping, the trion charge is assumed to be negative for the discussion below. Note that this assumption is not critical for ML WSe$_2$ as the spin-splitting of both the conduction band ($\approx30$~meV at $k=K_{\pm}$ \cite{Liu:2013a,Kosmider:2013a}) and valence band ($\approx430$~meV \cite{Zhu:2011a}) are substantial. The identification of the transitions is based on the polarization analysis shown in figures \ref{fig:fig1}c and d. Under linearly polarized laser excitation, only the highest energy peak shows linear polarization in emission and is therefore ascribed to the X$^0$, as a coherent superposition of valley states is created \cite{Jones:2013a}. 
This observation of exciton alignment is independent of the direction of the incident laser polarization, which confirms that the observed linear polarization is not due to macroscopic birefringence in the sample (see figure \ref{fig:figS6}). The strong remaining coherence in figure \ref{fig:fig1}d following non-resonant excitation hints at a direct optical generation of the neutral exciton X$^0$ for the laser energy used, energetically below the free carrier absorption and well below the B-exciton. 
Under circularly polarized excitation in figure \ref{fig:fig1}c, the two highest energy transitions are strongly polarized, as expected for the X$^0$ and the trion. The clear separation by 30 meV of the trion (PL FWHM 15~meV) and neutral exciton (PL FWHM 10~meV) is a major advantage compared to current MoS$_2$ ML samples for the independent investigation of the valley dynamics. Energetically below the trion emission we record two emission peaks that we assign to localized exciton complexes and that are accordingly labelled L1 and L2.\\
\indent In TRPL experiments we observe striking differences between the main transitions, as can be seen in figure \ref{fig:fig1}e and figure \ref{fig:fig2}. We first discuss the emission times, that can be compared in figure \ref{fig:fig2}. The main X$^0$ emission time cannot be resolved by our experiment, it decays within 4ps as shown in figure \ref{fig:fig2}, in a very similar way to ML MoS$_2$ \cite{Lagarde:2014a}. As a result of the short PL emission time, the coherence time could be as short as a few ps and still result in a strong linear polarization degree of the time-integrated PL \cite{Jones:2013a}. If the short X$^0$ emission time is limited by radiative recombination in this system with predicted exciton binding energies of several hundred meV \cite{Qiu:2013a,Ramasubramaniam:2012a} or by non-radiative processes, is still an open question. The weak PL emission at later times can be well fitted by a simple exponential decay with a characteristic time of 30 ps. The origin of this longer time may be attributed to exciton-phonon scattering, which scatters the exciton out of the light cone. The excitons have subsequently to reduce their momenta via scattering to return to the light cone to recombine radiatively \cite{Korn:2011a,Sanvitto:2000a}. Another possible origin of the emission comes from neutral excitons localised at fluctuations of the Coulomb potential \cite{Deveaud:1991a,Amand:1993a}. Note that this type of weak localisation is qualitatively different from forming bound states such as the $D^0X$ in GaAs, that might have parallels to the complexes L1 and L2. \\
\indent The trion PL emission can be fitted by a bi-exponential decay. We observe no measurable risetime of the trion PL signal within our resolution. The initial trion decay is clearly longer than for the X$^0$ as can be seen already in figure \ref{fig:fig1}e and we extract a decay time of about 18~ps, see figure \ref{fig:fig2}. At longer times we extract an emission with a decay of 30ps, just as in the case of the X$^0$. In general the longer emission times allow for a more detailed polarization analysis in the case of the trion compared to the ultra-short X$^0$ emission. The trion emission time being longer than the X$^0$ emission time is a trend also observed in III-V \cite{Finkelstein:1998a} and II-VI \cite{Vanelle:2000a} semiconductor quantum wells, where the longer trion emission time was ascribed to a lowering of the oscillator strength due to a stronger localisation.\\
\indent For the peak labelled L1 we record a clear, mono-exponential decay. The majority of photons resulting from L1 recombination are emitted after the main X$^0$ and trion recombination, see figures \ref{fig:fig1}e and \ref{fig:fig2}. The decay time extracted here is 30ps. The considerably less intense transition L2 also decays mono-exponentially, albeit with a considerably longer characteristic time of 80s. In temperature dependent measurements, we find at T=100~K that the L2 emission is negligible compared to L1, X$^0$ and the trion lines (see figure \ref{fig:figS3}). The emission times measured here for X$^0$, trion, L1 and L2 transitions remain essentially constant for the the laser excitation energies used ($1.851,1.893$ and $1.968$~eV), as documented in figure \ref{fig:figS4}.
\begin{figure}
\includegraphics[width=0.47\textwidth]{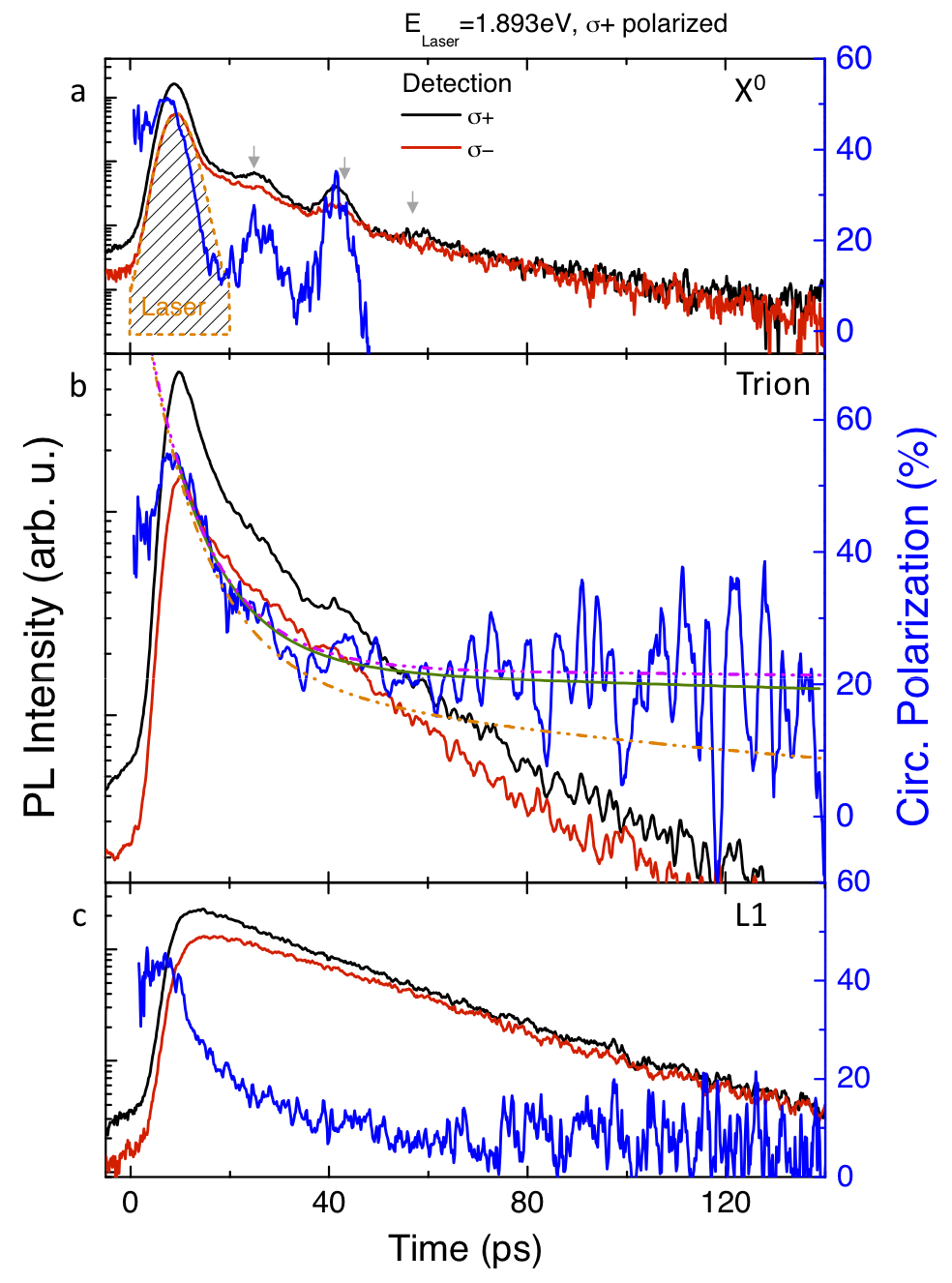}
\caption{\label{fig:fig3} \textbf{Time resolved photoluminescence.} T=4K. Laser polarization $\sigma^+$. (a) Left axis: X$^0$ PL emission (in log-scale) co-polarized (black) and cross-polarized (red) with respect to the excitation laser as a function of time. Right axis: circular polarization degree of the PL emission. As the PL intensity decays very quickly, we clearly observe a periodic signal of laser reflections, marked by arrows (b) same as (a), but for trion emission. The polarization well reproduced by a bi-exponential decay using an initial, fast decay time of 12~ps and a long decay time 1~ns (solid green line). Lower  bounds (dotted orange line for 150ps) and upper bounds  (dotted purple line for 3~ns) of the slow decay are shown. (c) same as (a), but for localized exciton emission L1.}
\end{figure} 

\indent We now discuss the time evolution of the PL polarization that gives access to the valley dynamics.  Due to the fast decay time of the main X$^0$ emission (limited by our temporal resolution), we cannot extract a meaningful polarization decay time.  According to recent estimations \cite{Yu:2014a,Yu:2014b} neutral excitons from non-equivalent K valleys in TMDCs are coupled by the Coulomb exchange interaction. This could lead to a rapid decay of the optically initialized circular polarization, in close analogy to neutral exciton depolarization in GaAs quantum wells \cite{Maialle:1993a} and will contribute to the fast inter-valley relaxation decay observed in 1ML MoS$_2$ in pump-probe experiments \cite{Mai:2014a,Wang:2013d}. The coupling of a neutral exciton created with $\sigma+$ light in the K$_+$ valley to an exciton in the K$_-$ valley is efficient as it does \underline{not} rely on single carrier (electron or hole) spin flips, which are energetically forbidden.\\
\indent The initial trion emission time is much longer than the one of the X$^0$ in figures \ref{fig:fig2} and \ref{fig:fig3}b, it decays within a few tens of ps, well above the temporal resolution of our experiment. The longer trion emission time allows us to access the time evolution of the valley polarization: We first observe an initial decay from 50\% down to about 20\% with a characteristic time of 12ps. Concerning the origin of this initial decay: The trion could be either generated directly following photon absorption (phonon assisted process), or by a localised electron capturing a free exciton. This second scenario is unlikely, as within the initial decay of 12ps the neutral exciton polarization would have already decayed to zero before capture, which is in contradiction with the strong, remaining valley polarization. For times $t>\approx40~ps$ we can infer a typical decay time of about 1ns for this substantial polarization, as indicated in figure \ref{fig:fig3}b. Please note that a typical decay time of 150~ps gives an estimation that is clearly below the observed experimental polarization, whereas a decay time of 3~ns still gives a good fit (see dotted lines in figure \ref{fig:fig3}b). The trion PL polarization dynamics show clear, experimental proof of the robustness of the optically initialised valley polarization. 
Measurements carried out at different laser excitation energies show similarly encouraging results (see figure \ref{fig:figS5}). This is in contrast to ML MoS$_2$, which shows a strong decrease of the PL polarization when the laser excitation energy increases \cite{Kioseoglou:2012a,Lagarde:2014a}. 
This difference could be due to the fact that the $\Gamma$ valence states are very close in energy to the K states  (a few meV \cite{Kormanyos:2013a}) in MoS$_2$ whereas for WSe$_2$ the splitting energy is considerably larger.\\
\indent The emission we labelled L1 shows a fast initial polarization decay with a characteristic time of about 13~ps before reaching a polarization plateau as in the trion case, albeit at a smaller value around 8\%. In reference \cite{Jones:2013a} in a gated WSe$_2$ ML structure a peak approximately at the energy of L1 was ascribed to a fine structure split trion state. Although the linewidth of the PL emission of the X$^0$, trion and L1 is considerably smaller than in MoS$_2$, we cannot extract any information at this stage concerning an eventual fine structure due to strong Coulomb exchange in this system \cite{Yu:2014b}, which might influence the observed polarization dynamics.\\
\indent In conclusion, time resolved PL experiments in monolayer WSe$_2$ allow to access directly the valley dynamics by monitoring the charged exciton (trion) emission. We measure a valley polarization decay time longer than 1 ns. To verify valley stability beyond this time range pump-probe measurements are needed. Similar results can be expected for ML MoS$_2$, but are currently far more difficult to extract due to the broader (roughly $3.5\times$) linewidth at 4K and also 300K in exfoliated samples.

\section{Acknowledgements}
\label{sec:Acknowledgements}
We acknowledge partial funding from ERC Starting Grant No. 306719 and Programme Investissements d'Avenir ANR-11-IDEX-0002-02, reference ANR-10-LABX-0037-NEXT. We thank I. Gerber and M.M. Glazov for stimulating discussions.

\begin{figure}
\includegraphics[width=0.47\textwidth]{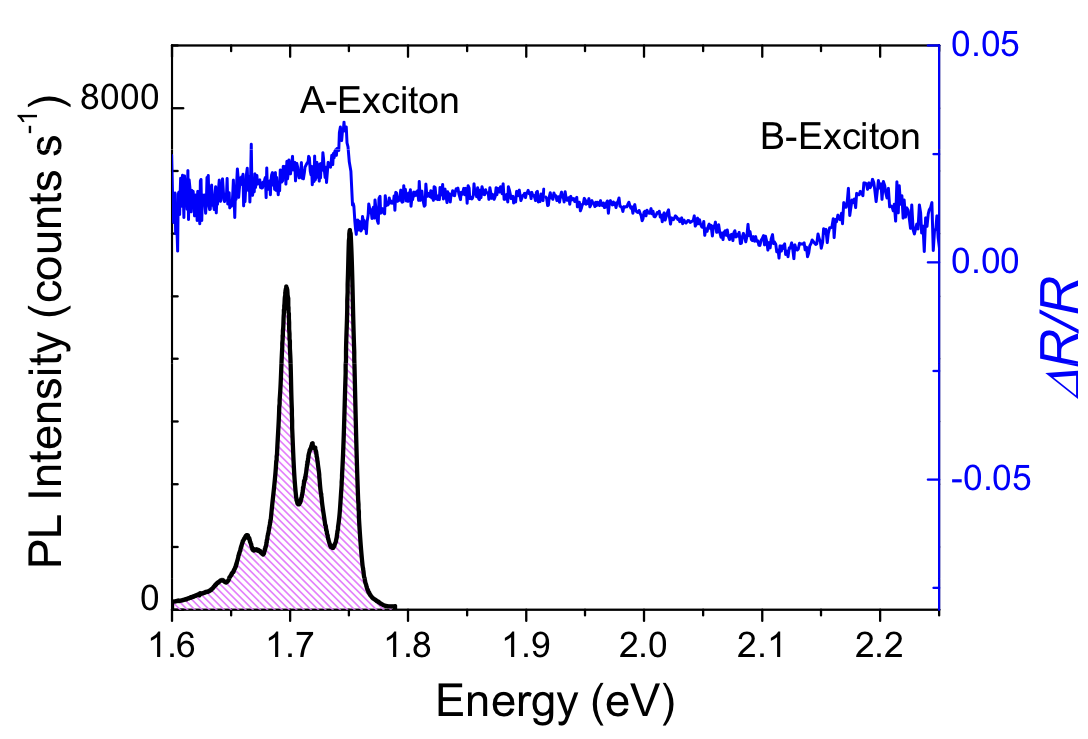}
\caption{\label{fig:figS1}  T=4K. The position of the neutral A-exciton emission (blue curve) is confirmed in reflectivity measurements (blue curve) compared to the PL emission of the A-exciton (black curve). We measure an energy difference between A- and B- exciton of  $410\pm15~$meV, in good agreement with theory \cite{Zhu:2011a} and values extracted from hot PL emission \cite{Zeng:2013a}.}
\end{figure} 

\begin{figure}
\includegraphics[width=0.47\textwidth]{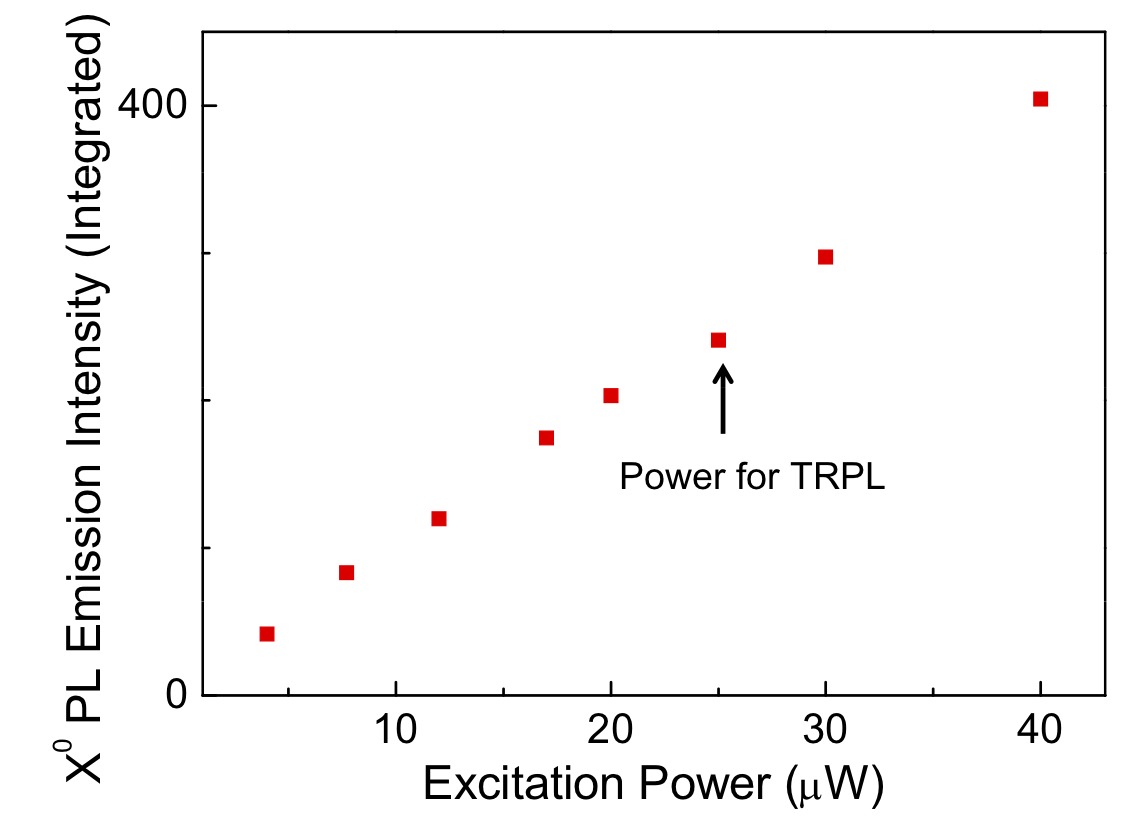}
\caption{\label{fig:figS2}  T=4K. The time resolved measurements were carried out at low power well within the linear absorption regime. The laser power was kept low in all experiments (i) to avoid sample damage, a problem studied in detail in Raman experiments \cite{Li:2013a} (ii) to avoid secondary emission from high signal levels of the phosphorus screen of the streak camera.}
\end{figure} 

\begin{figure}
\includegraphics[width=0.47\textwidth]{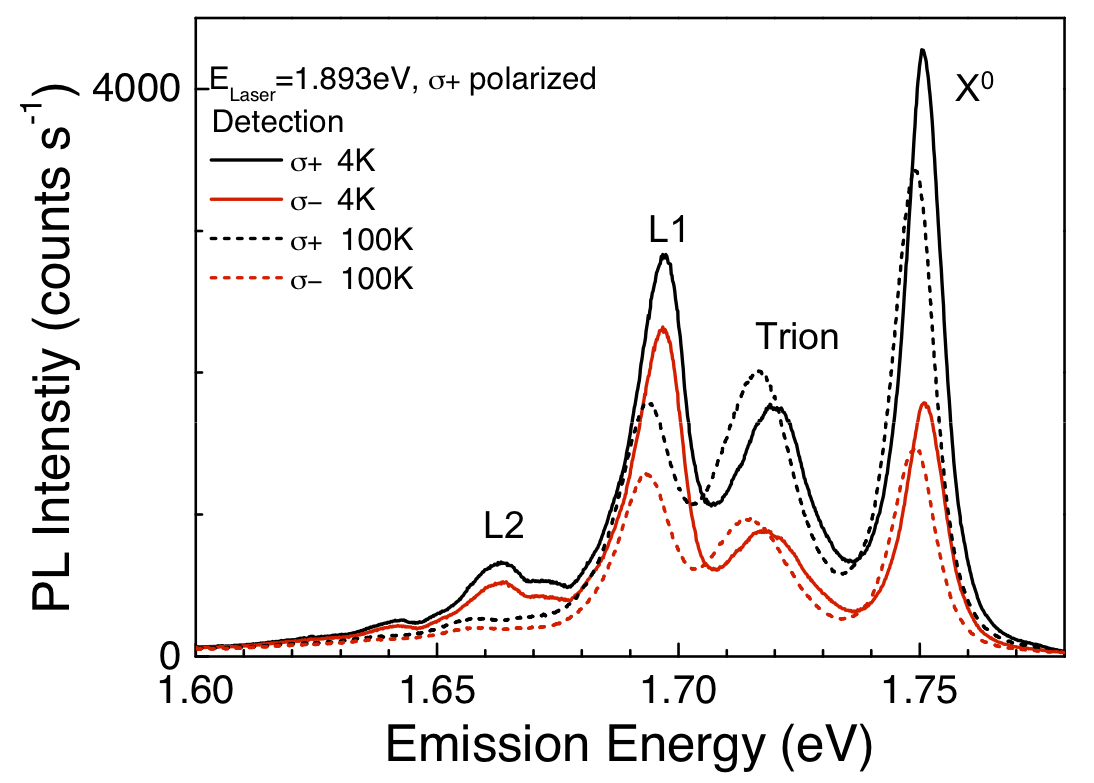}
\caption{\label{fig:figS3}  The PL intensity of the peak L2 is strongly reduced when comparing the spectra at 4K and 100K, confirming the localised nature of the emission. }
\end{figure} 

\begin{figure}
\includegraphics[width=0.47\textwidth]{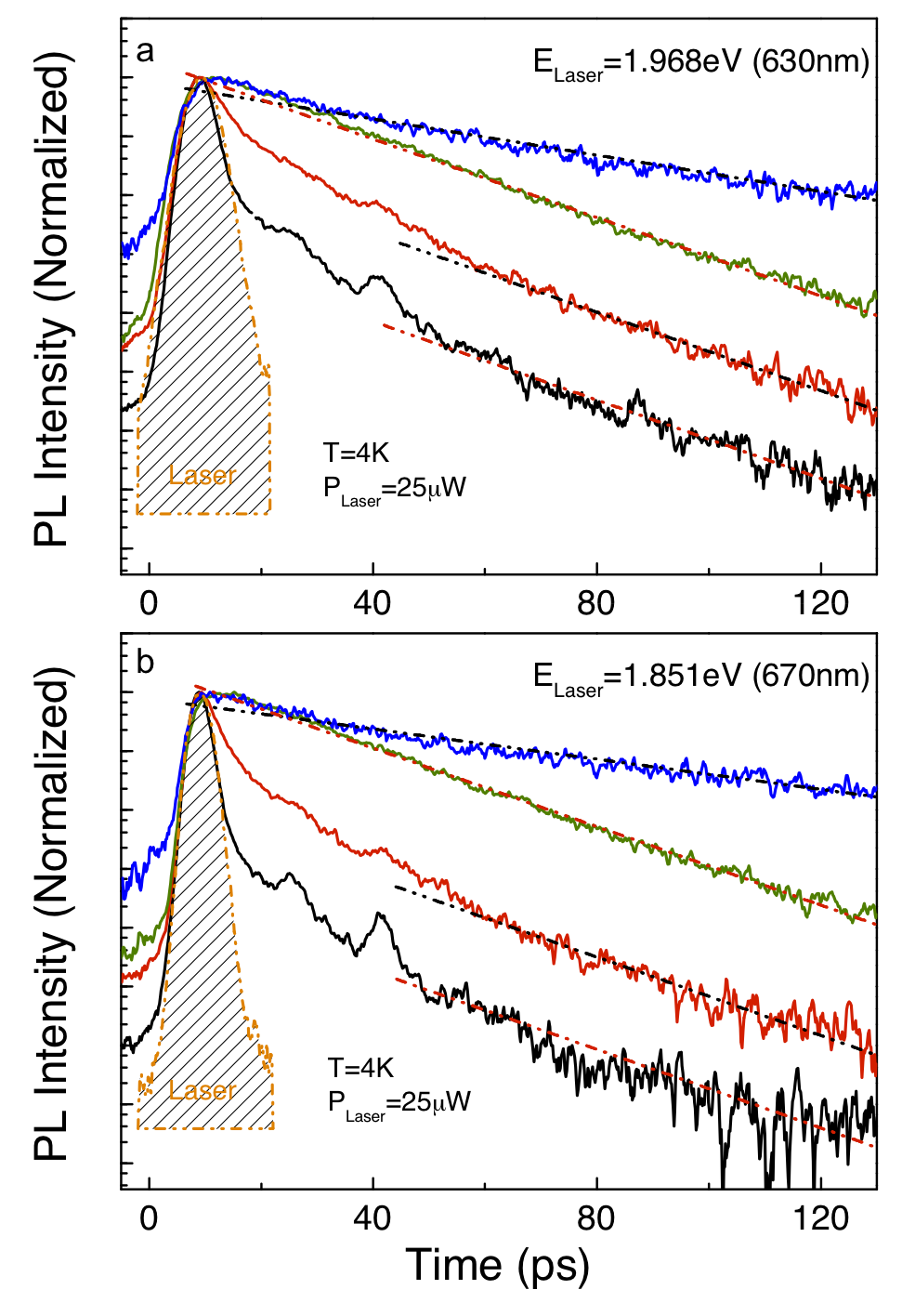}
\caption{\label{fig:figS4} \textbf{Emission time for different excitation laser energies.} The PL emission is shown for two additional laser excitation energies E$_\text{Laser}$ as compared to the main text. The color code is the same as in the main text: X$^0$ - black, Trion - red, L1 - green and L2 - blue. The decay times (fits indicated by dotted lines) are the same as for E$_\text{Laser}=1.893$~eV used in the main text for X$^0$, Trion and L1. The L2 decay time in (a) is 65~ps and in (b) 78~ps.}
\end{figure} 

\begin{figure}
\includegraphics[width=0.47\textwidth]{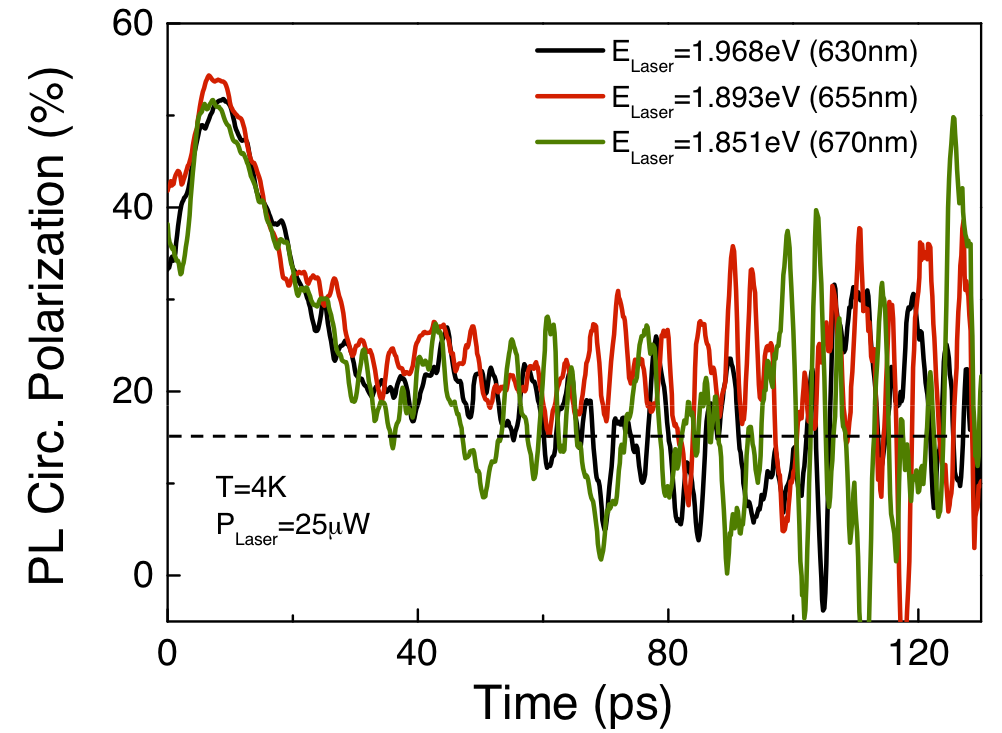}
\caption{\label{fig:figS5}  \textbf{Valley dynamics monitored through trion emission for different excitation laser energies.} The time evolution of the PL polarization i.e. the valley dynamics is plotted for the trion using three different excitation laser energies. In all three cases the dynamics is very similar, with a long lasting, stable polarization after an initial, fast decay. The dashed line is a guide to the eye.}
\end{figure} 

\begin{figure}
\includegraphics[width=0.47\textwidth]{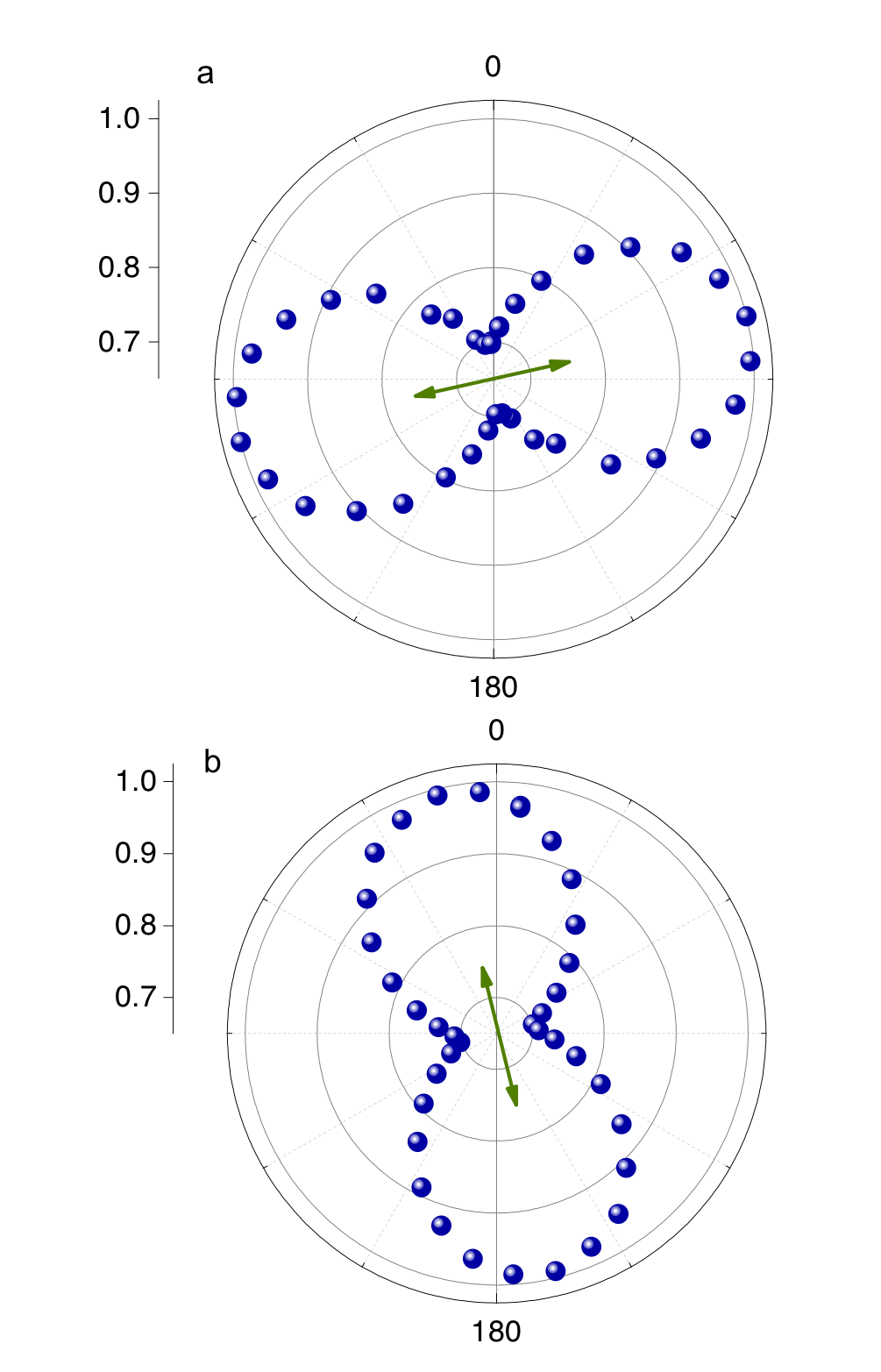}
\caption{\label{fig:figS6}  \textbf{Valley coherence of neutral exciton X$^0$.} T=4K. The excitation laser polarization is linear. Here the time integrated PL emission intensity of the X$^0$ is plotted as a function of the angle of the linear polariser in the detection path for different linear polarization planes of the excitation laser (a) and (b). The X$^0$ polarization plane follows the excitation laser and is not linked to a specific lattice axis or symmetry, which confirms the generation of neutral exciton valley coherence \cite{Jones:2013a}}
\end{figure} 

\end{document}